
\input harvmac
\newcount\figno
\figno=0
\def\fig#1#2#3{
\par\begingroup\parindent=0pt\leftskip=1cm\rightskip=1cm\parindent=0pt
\baselineskip=11pt
\global\advance\figno by 1
\midinsert
\epsfxsize=#3
\centerline{\epsfbox{#2}}
\vskip 12pt
{\bf Fig. \the\figno:} #1\par
\endinsert\endgroup\par
}
\def\figlabel#1{\xdef#1{\the\figno}}
\def\encadremath#1{\vbox{\hrule\hbox{\vrule\kern8pt\vbox{\kern8pt
\hbox{$\displaystyle #1$}\kern8pt}
\kern8pt\vrule}\hrule}}

\overfullrule=0pt

%

%

\font\zfont = cmss10 

\def\bigone{\hbox{1\kern -.23em {\rm l}}}
\def\ZZ{\hbox{\zfont Z\kern-.4emZ}}

\Title{hep-th/9409111, IASSNS-HEP-94-72}
{\vbox{\centerline{IS SUPERSYMMETRY REALLY BROKEN?}}}
\smallskip
\centerline{Edward Witten\footnote{*}{Research supported by NSF grant
\#PHY92-45317}}
\smallskip
\centerline{\it School of Natural Sciences, Institute for Advanced Study}
\centerline{\it Olden Lane, Princeton, NJ 08540, USA}\bigskip
\baselineskip 18pt

\medskip

\noindent
In $2+1$ dimensions, in the presence of gravity, supersymmetry can
ensure the vanishing of the cosmological constant without requiring
the equality of bose and fermi masses.

\Date{October, 1994}

Within the known structure of physics, supergravity  in four dimensions
leads to a dichotomy: either the symmetry is unbroken and bosons
and fermions are degenerate, or the symmetry is broken
and the vanishing of the cosmological constant is difficult to understand.
  Neither of these alternatives appears satisfactory.

Even when supersymmetry is unbroken, the cosmological constant might
not vanish, but with suitable reasonable assumptions -- such as an $R$
symmetry or an underlying string theory --
it vanishes naturally.

The purpose of this short note is to point out that in $2+1$ dimensions,
the unsatisfactory dichotomy does not arise: supersymmetry can
explain the vanishing of the cosmological constant without leading
to equality of bose and fermi masses.   Perhaps the same would be true
in a suitable modified framework of physics in $3+1$ dimensions.

First of all, to get a $2+1$ dimensional supergravity theory
with unbroken supersymmetry in which the cosmological
constant naturally vanishes, one can, for instance, take
the dimensional reduction of any $3+1$ dimensional theory with
this property.  Do bosons and fermions have equal masses in such a theory?
To prove that they do, one must use the unbroken global supercharges.
In local supersymmetry, there are ${\it a priori}$ infinitely many
supercharges (the generator being an arbitrary spinor field), but
the spontaneously broken ones are not useful for controlling the spectrum.

The unbroken supercharges are determined by spinor fields that are
covariantly constant at infinity; in fact, the space of unbroken
supercharges can be identified naturally with the space of possible
asymptotic values of a spinor field that is asymptotically covariantly
constant.

Now we meet the fact that, in $2+1$ dimensions (in a world with vanishing
cosmological constant), any state of non-zero energy produces a geometry
that is asymptotically conical.  In such a conical geometry, there
are no covariantly constant spinors, so there are no unbroken
supersymmetries.  Thus, in $2+1$ dimensional supergravity,
even when supersymmetry
applies to the vacuum and ensures the vanishing of the vacuum energy,
it does not apply
to the excited states.

Of course, this detailed mechanism is special to $2+1$ dimensions,
but it at least makes one wonder whether, in a suitable framework,
there can be an analogous phenomenon in $3+1$ dimensions.

I would like to acknowledge the hospitality of the Santa Fe Institute
where a version of this argument was presented some years ago.
\end